\documentclass{revtex4}

\usepackage{graphicx}
\begin{document}
\bibliographystyle{revtex}

\hfill{OUNP-2001-04}

\hfill{December 2001}


\title{R\&D for a CCD Vertex Detector for the High Energy Linear $e^+e^-$ Collider}



\author{P.N. Burrows}
\email[]{p.burrows@physics.ox.ac.uk}
\affiliation{Oxford University}

\author{(Representing the LCFI (UK) Collaboration)}
\noaffiliation

\date{\today}

\begin{abstract}
I summarise the status of the LCFI Collaboration R\&D programme for a 
CCD-based vertex detector for the linear collider.
\end{abstract}

\maketitle

\section{Introduction}

Charge-coupled devices (CCDs) were originally applied in high-energy particle 
physics at a fixed-target charm-production experiment, and their utility
for high-precision vertexing of short-lived particles was quickly 
realised~\cite{accmor}. More recently two generations of CCD vertex detectors
(VXDs) were used in the $e^+e^-$ colliding-beam environment of the SLD experiment
at the first linear collider, SLC, at SLAC.

CCDs are silicon pixel devices which are widely used for imaging; one common
application is in home video cameras, and there is extensive industrial
manufacturing experience in Europe, Japan and the US. 
CCDs can be made with high pixel granularity. For example, those used at SLD
comprise 20$\times$20 $\mu$m$^2$ pixels, offering the possibility of intrinsic
space-point resolution of better than 3 $\mu$m, determined from the centroid of 
the small
number of pixels which are hit when a charged particle traverses the device.
The active depth in the silicon is only 20 $\mu$m, so each pixel is
effectively a cube of side 20 $\mu$m, yielding true 3-dimensional spatial
information. Furthermore, this small active depth allows CCDs to be made
very thin, ultimately perhaps as thin as 20 $\mu$m, which corresponds to significantly
less than 0.1\% radiation length ($X_0$), and yields a 
very small multiple scattering of charged particles. 
Also, large-area CCDs can be made for scientific purposes, allowing an
elegant VXD geometry with no azimuthal gaps or dead-zones for readout
cables or support structures. 

The combination of superb spatial resolution, low multiple scattering,
and large-area devices, with a decade of operating experience
at SLC, makes CCDs a very attractive
option for use in a vertex detector at the next-generation linear
collider (LC). 

\section{SLD CCD VXD Experience}

The SLD experiment has utilised three CCD arrays for heavy-flavour
tagging in $Z^0$ decays. In 1991 a 3-ladder prototype detector, VXD1,
was installed for initial operating experience. In 1992 a complete
four-layer detector, VXD2, was installed and operated until 1995. 
VXD2~\cite{vxd2} utilised 64 ladders arranged in 4 incomplete azimuthal
layers, with $\geq$2-hit coverage extending down to polar angles within
$|\cos\theta|\leq0.75$. The device contained 512 roughly
$1\times1$cm$^2$ CCDs, giving a total of 120M pixels.

In 1996 a brand new detector, VXD3~\cite{vxd3}, was installed.
The main improvement was to utilise much larger, $8\times1.6$cm$^2$,
and thinner ($\times$ 3)  CCDs,
which allowed significantly improved azimuthal- and polar-angle coverage.
Ladders were formed from two CCDs placed end-to-end (with a small
overlap in coverage) on opposite sides of a beryllium support beam,
and arranged in 3 complete azimuthal layers, with a `shingled'
geometry to ensure no gaps in azimuth.
96 CCDS were used, giving a total count of 307M pixels.

In operations from 1996 through 1998 VXD3 performed beautifully,
yielding a measured single-hit resolution of $\sim$3 $\mu$m, and a track
impact-parameter resolution of 8 $\mu$m (10 $\mu$m) in $r-\phi$
($r-z$) respectively, measured using 46 GeV $\mu$ tracks in $Z^0$ $\rightarrow$
$\mu^+\mu^-$ events. The multiple scattering term was 
$33/p\sin^{3/2}\theta$ $\mu$m. 
For inclusive $b$-hemisphere tagging a sample purity of 98\% was
obtained with a tag efficiency of over 60\%, and for inclusive 
$c$-tagging a sample purity of around 80\% was obtained with a
tag efficiency of about 20\%. This performance is `state-of-the-art' today.

\section{Linear Collider Physics Demands}

The LC will probably be built to operate
at c.m. energies in the range between the LEP2
energy of around 200 GeV and up to 0.8 or 1 TeV. 
Many of the interesting physics processes can be
characterised as multijet final states containing heavy-flavour
jets. 
It should be noted that charm- and $\tau$-tagging, as well as $b$-tagging,
will be very important. For example, measurements of the branching ratios
for (the) Higgs boson(s) to decay into $b$, $c$, and $\tau$ pairs 
(and/or $W$, $Z^0$ and $t$ pairs for a heavy Higgs)
will be
crucial to map out the mass-dependence of the Higgs coupling and to
determine the nature (SM, MSSM, SUGRA $\ldots$) of the Higgs particle(s).
Because of this multijet structure, even at $\sqrt{s}$ = 1 TeV many of these
processes will have jet energies in the range 50 $\rightarrow$ 200 GeV, which is not
significantly larger than at SLC, LEP or LEP2. The track momenta will be
correspondingly low. For example, at $\sqrt{s}$ = 500 GeV the mean track
momentum in $e^+e^-$ $\rightarrow$ $q\overline{q}$ 
events is expected to be around 2 GeV/$c$. Hence, for the majority of tracks,
multiple scattering in thick detectors will limit 
the impact-parameter resolution, as was the case even with SLD VXD3, and will
compromise the flavour-tagging performance, most seriously for charm and $\tau$. 

Furthermore, some of these processes may lie close to
the boundary of the accessible phase space, suggesting that extremely high
flavour-tagging efficiency will be crucial for identifying a potentially small
sample of events above a large multijet combinatorial background.
It is worth bearing in mind that a doubling of the single-jet
tagging efficiency at high
purity is equivalent to a luminosity gain of a factor of 16 for a 4-jet
tag; it is likely to be a lot cheaper (and easier) 
to achieve this gain by
building a superior VXD than by increasing the luminosity of the
accelerator by over an order of magnitude.
 
\section{LC VXD R\&D Programme}

\begin{figure}
 \includegraphics[width=80mm]{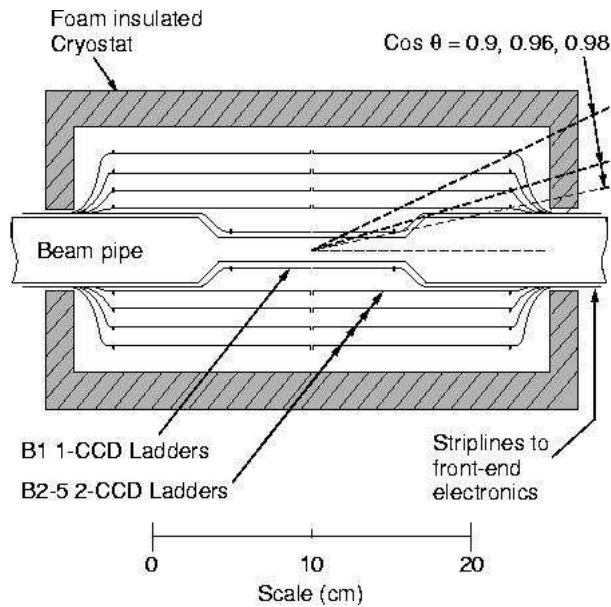}
 \caption{Schematic of the LC CCD vertex detector.}
 \label{e3032fig1}
 \end{figure}

The LCFI VXD conceptual design is illustrated in Fig.~\ref{e3032fig1}.
Simulation of the flavour-tagging performance is described elsewhere~\cite{stefania}.
Table~\ref{factors} summarises the improvement factors that it is
hoped to achieve, relative to the current SLD VXD3, for various key
parameters; the most challenging are the ladder thickness and readout speed.

\begin{table}
\begin{center}
\begin{tabular}{|l|c|c|c|}
\hline
Item &             SLD & LC  & factor \\ \hline
 & & &  \\
longest CCD (mm) & 80  & 125 &   1.6 \\  
largest CCD area (mm$^2$) &  1280  &  3000 &  2.3 \\  
\# ladders          & 48  &  64 &   1.3 \\  
\# pixels (M)       & 307  & 900 &   2.3 \\  \hline
ladder thickness (\% $X_0$)  &  0.4 &  0.06 &  7 \\  
pixel readout rate (MHz)  & 5  & 50 &   10 \\  
column-parallel r/o & $\times$ & $\sqrt{}$ & 1000 \\
\hline
\end{tabular}
\caption{CCD performance improvement factors required for the LC VXD}
\label{factors}
\end{center}
\end{table}

\subsection{Ladder Thickness}

An aggressive option of `unsupported' silicon is being pursued. In this
mode the CCDs would be back-thinned to 50 or 60$\mu$m (0.06\% $X_0$), assembled into
a ladder structure, and held under 
tension at the ends. Although the production of CCDs as thin as 20 $\mu$m has 
been achieved for use in astronomy, the use of such `unsupported' devices has
not been tried before, and a number of key issues are being addressed.
Of primary concern is the degree of mechanical stability achieveable. This
has both `static' and `dynamic' aspects. We require shape reproducibility at
the level of a few microns under temperature cycling between room temperature
and about 180K. We also require insensitivity to vibrations, eg. possible
flow-driven oscillations caused by the coolant N$_2$ gas. 

A test rig has been assembled and used to investigate the stability of
thin prototype ladders comprising two dummy CCDs glued together to form
25cm-long structures. The ladder is pinned to the support jig at one end, 
and tensioned via a spring mechanism, with 
a sliding joint to the jig, at the other. Tests have been performed using glass and
unprocessed silicon of 60 $\mu$m thickness. The results are extremely
encouraging. For modest tensions, above 15N, the position reproducibility
is better than 3 $\mu$m under successive relaxation and reapplication of
the tensioning mechanism. The stability under temperature cycling has also
been investigated. Here the type of adhesive, the relative CTEs of the
glue, ladder and supporting blocks, the pattern of application of
the glue, and the alignment of the blocks w.r.t. the ladder are all
crucial elements that have been studied using both prototypes and FEA
simulations. An optimal solution has been found, which will now be tried with
actual thin CCDs. 

Metrology apparatus for these stability tests has been set up at RAL 
and Oxford. The RAL system uses a commercial microscope CMM that is
excellent for single-point measurements. The Oxford system is a 
custom-made white-light interferometer that allows micron-level complete surface
profile measurements to be made of ladders as large as 30 $\times$ 2.5 cm$^2$. 

A possible fall-back option is to support the thin CCDs on a thin
flat Be beam with an intrinsic `omega' or `V'
shape for rigidity. Finite-element analysis simulations have shown that
such structures offer the possibility of small, and predictable,
deformations under temperature cycling of the order of tens of $\mu$m.
If the support beam comprises 250 $\mu$m Be-equivalent, or 0.07\% $X_0$,
and the adhesive an additional 0.02\% $X_0$, the CCDs could be fully thinned
to 20 $\mu$m, yielding a total ladder material
budget as low as 0.11\% $X_0$. The handling and assembly of such thin devices
will be addressed if the `unsupported' option proves untenable.

\subsection{Readout Speed} 

The allowed radial position of the innermost layer w.r.t. the beamline is
strongly influenced by the accelerator-related backgrounds, and is
correlated with the pixel readout rate, which determines the 
hit density accumulated during the CCD readout cycle, and hence the
degree of fake hit confusion for bona fide tracks.
The main sources of accelerator-related backgrounds are:
muons from beam interactions with upstream collimators;
$e^+e^-$ pairs from converted photons and `beamstrahlung';
photoproduced neutrons from the interaction region
material and back-shine from the beam dumps;
hadrons from beam-gas and $\gamma\gamma$ interactions.

From the VXD occupancy point-of-view the most serious are the
$e^+e^-$ pairs. For example, beam-beam interaction simulations
indicate that tens of thousands of pairs will be created {\it per
bunch crossing} of the beams. 
For a VXD layer-1 radius of around 13mm, 
a large detector solenoidal magnetic field will be required to contain the bulk
of the pairs within the beampipe, and maintain an acceptably low hit density.
Field strengths of between 3 and 6 T are being considered
by the detector working groups. At both NLC (6 T) and TESLA (4 T) roughly 
0.03 hits/mm$^2$/bunch crossing are expected in VXD layer 1.
This translates to rougly 6 hits/mm$^2$/bunch {\it train} at NLC, and
roughly 90 hits/mm$^2$/bunch {\it train} at TESLA. Since the pixel density
is 2500/mm$^2$, a readout that integrates over one complete train is
acceptable for NLC, but would yield an uncomfortable 4\% occupancy at TESLA.
This is not disastrous, but studies show~\cite{stefania} that some
pollution of tracks with background hits would result in this crucial 
layer, closest to the IP.

The requirements are therefore to achieve a complete detector readout between NLC 
bunch trains, i.e. within about 8ms, and to read out roughly 10 times per train
at TESLA, i.e. within about 100 $\mu$s. The NLC goal can be met with a
factor of 10 increase in pixel-readout rate relative to what was achieved
at SLD, namely 50 MHz. The TESLA goal requires, in addition,
parallelisation of the CCD readout; we are investigating the design of
a CCD in which every column is read out through an independent output node.
This will require the output-node pads and associated readout electronics
to be laid out on a pitch of 20 $\mu$m. This is challenging, but preliminary
design work for an output circuit on this pitch has been done, and at least
one company is able to produce ADCs on the same pitch.

We have outlined a staged approach for developing a column-parallel
CCD with the required pizel readout speed, starting at 0.5 MHz, progressing
to 5 MHz, and hopefully reaching 50 MHz. The design work, in collaboration
with Marconi Applied Technologies, is well advanced. In addition we are
bench-testing a standard CCD that has the promise of reaching 70 MHz serial 
readout speed; this chip has so far been driven at 10 MHz with good signal/noise
performance. VME-based drive and readout electronics for 50 MHz operation 
are under construction at RAL. 

\subsection{Radiation Damage Studies} 

The neutron flux in the inner detector is expected to be at the level of 
10$^9$/cm$^2$/year. This is about an order of magnitude below the threshold at which
a non-negligible charge transfer inefficiency (of order $10^{-4})$
results from charge-trapping by
damage centres. A number of promising ideas offer the possibility of further headroom.
For example, lower-temperature operation may increase the tolerance via trap
`freezeout'. Trap filling via auxilliary charge injection is another possibility.
These ideas warrant further investigation, and low-temperature studies are 
expected to be performed at the Liverpool test setup. 

\section{Summary and Outlook}

CCDs offer a very attractive option for a high-energy
linear collider vertex detector. CCD VXDs have been `combat-tested'
at the first linear collider, SLC, and have allowed SLD to achieve
unrivalled $b$ and $c$-jet tagging performance.
We are addressing a number of R\&D issues to permit successful 
application of this technology at the LC. There is considerable interest and overlap
with other scientific communities, for example astronomy, remote sensing and
X-ray imaging.


\end{document}